\documentclass[a4paper,11pt]{article}

\usepackage[utf8]{inputenc}
\usepackage[T1,T2A]{fontenc}
\usepackage{amsmath,amsfonts,amssymb}

 \usepackage{graphicx}
\usepackage{wrapfig}
\usepackage{subfig}
\usepackage{hyperref}
\usepackage[dvipsnames]{xcolor}

\begin{document}

\begin{titlepage}

\begin{center}
{\LARGE {\bf Higgs Portal Dark Matter Freeze-in  at Stronger Coupling: Observational Benchmarks}} \\
\vspace{2cm}
{\bf Giorgio Arcadi$^{\,1,2}$, Francesco Costa$^{\,3}$, Andreas Goudelis$^{\,4}$,
Oleg Lebedev$^{\,5}$}
\end{center}

\begin{center}
  \vspace*{0.25cm}
  $^1$ \it{Dipartimento di Scienze Matematiche e Informatiche, Scienze Fisiche e Scienze della Terra,\\ Universita
degli Studi di Messina, Via Ferdinando Stagno d'Alcontres 31, I-98166 Messina, Italy} \\
 \vspace{0.2cm}
      $^2$ \it{INFN Sezione di Catania, Via Santa Sofia 64, I-95123 Catania, Italy} \\
      \vspace{0.2cm}
      $^3$ \it{Institute for Theoretical Physics, Georg-August University G\"ottingen, \\
  Friedrich-Hund-Platz 1, G\"ottingen D-37077, Germany} \\
      \vspace{0.2cm}
      $^4$ \it{Laboratoire de Physique de Clermont Auvergne (UMR 6533), CNRS/IN2P3, Univ.\ Clermont Auvergne, 4~Av.\ Blaise Pascal, F-63178 Aubi\`ere Cedex, France} \\
      \vspace{0.2cm}
 $^5$ \it{Department of Physics and Helsinki Institute of Physics,\\
  Gustaf H\"allstr\"omin katu 2a, FI-00014 Helsinki, Finland}\\
\end{center}

\vspace{2.5cm}

\begin{center} {\bf Abstract} \end{center}
\noindent
We study freeze-in production of Higgs portal dark matter (DM)  at temperatures far below the dark matter mass. 
The temperature of the Standard Model  (SM) thermal bath may have never been high such that dark matter production
via thermal emission has been  Boltzmann-suppressed. This allows for a significant coupling between the Higgs field and DM,
which is being probed by the direct DM detection experiments and invisible Higgs decay searches at the LHC. 
We delineate the corresponding parameter space in the Higgs portal framework with dark matter of spin 0, 1/2 and 1.
\end{titlepage}


\tableofcontents

\vspace{1cm}

\section{Introduction}

The dark matter (DM)  component of the Universe may be due to a new particle which has no Standard Model (SM) quantum numbers. 
Interactions of such a particle with the SM fields must be sufficiently weak or have a special structure to evade the strong direct DM detection bounds \cite{LZ:2022lsv}. This constraint appears to be difficult to
satisfy if DM is thermal such that its abundance is dictated by the thermal annihilation cross section into the lighter SM states. On the other hand, non-thermal dark matter 
does not suffer from this problem since its abundance is determined by a different mechanism. 

One of the most common non-thermal DM scenarios is known as ``freeze-in''  \cite{Hall:2009bx}  dark matter. In this case, it is produced by the SM thermal bath via  a feeble coupling to DM
such that the latter never thermalizes \cite{Dodelson:1993je}. The smallness of the coupling, on the other hand, makes this scenario difficult to test, if possible at all.
Furthermore, the model makes a strong and often unjustified assumption that the pre-existing DM abundance is zero.
This is problematic in standard cosmological settings \cite{Lebedev:2022cic}, where the thermal phase is preceded by a (high scale) inflation, preheating, etc. During these periods,  one expects violent particle production
both due to classical gravitational effects \cite{Ford:2021syk} and quantum-gravity generated interactions \cite{Lebedev:2022ljz}. In particular, gravity induces effective Planck-suppressed couplings between the inflaton field   $\phi$ and dark matter, e.g.
\begin{equation}
{1\over M_{\rm Pl}^2}\, \phi^4 s^2 ~,~ {1\over M_{\rm Pl}}\, \phi^2 \bar \chi \chi ~,...
\end{equation} 
Here $s$ and $\chi$ are spin 0 and spin 1/2 stable particles, respectively.
Such interactions are especially effective during the inflaton oscillation period and the
 resulting DM abundance normally exceeds the observed value by orders of magnitude, unless the quantum gravity effects are happen to be tiny or DM is extremely light \cite{Lebedev:2022cic,Koutroulis:2023fgp}.
 In the absence of full control over quantum-gravitational effects, this clearly presents a problem.
 
 One possibility to address this issue is to ``dilute'' the dark relics before reheating \cite{Lebedev:2022cic}. If the Universe  is dominated by a non-relativistic inflaton, the contribution of the relativistic species to the 
 energy density  decreases over time. Therefore, the relics produced right after inflation would be diluted if reheating occurs at late times corresponding to a low reheating temperature $T_R$. 
 In this case, the DM abundance is determined by the late-time dynamics and 
 the uncertainties associated with quantum gravity make no impact.
 
 These considerations motivate the possibility that the dark matter mass scale may be above the temperature of the SM bath \cite{Cosme:2023xpa}.
 In fact, there is no evidence that the temperature has ever been high, above 4 MeV \cite{Hannestad:2004px}. 
 If dark matter is significantly heavier than the maximal temperature $T_{\rm max}$, its thermal production would be Boltzmann-suppressed. 
  An interesting consequence of this possibility is that, even though DM is non-thermal, its coupling to the SM particles can be order one.
 As a result, it can be probed both via direct detection experiments and collider searches.
 
 The purpose of this work is to study the phenomenology of the simplest DM extensions of the SM, known as the ``Higgs portal'' \cite{Patt:2006fw}, in the regime of freeze-in at stronger coupling.      
 We consider singlet dark matter of spin 0, 1/2, 1 coupled to the SM only through the Higgs field \cite{Kanemura:2010sh,Djouadi:2011aa} and study the DM relic abundance 
 as a function of $T_R$ in the mass range from 1 GeV to 10 TeV, as well as the relevant direct/indirect detection and collider probes.

\subsection{Set-up}

In this work, we study dark matter freeze-in in the Higgs portal framework (see \cite{Lebedev:2021xey} for a review). That is, we assume that DM has no SM quantum numbers and communicates with the observable sector
via the Higgs coupling.
 The lowest dimension  interactions between the Standard Model fields  and the   singlets  $s, \chi, V_\mu$ of spin 0, 1/2, 1, respectively, 
 are given by 
  \begin{eqnarray}
 &&-\Delta {\cal L}_{\rm scal} = {1\over 2} \, \lambda_{hs} H^\dagger H \, s^2 \; , \label{h-s} \\
 &&-\Delta {\cal L}_{\rm ferm} = {1\over  \Lambda}\,    H^\dagger H  \,  \bar\chi \chi  +    {1\over  \Lambda_5}\,    H^\dagger H  \,     \bar\chi i\gamma_5 \chi  
  \label{h-chi}\; ,\\
&&-\Delta {\cal L}_{\rm vect} = {1\over 2}\, \lambda_{ hv } H^\dagger H \, V_\mu V^\mu \; , \label{h-v} 
 \end{eqnarray}
 where $s$ and $V_\mu$ are real, $\chi$ is  a $Majorana$ fermion;  $ \lambda_{hs}, \lambda_{ hv }$ are real dimensionless couplings; $\Lambda, \Lambda_5$ are   scales, and  
 we have allowed for CP violation in the fermion interactions \cite{Lopez-Honorez:2012tov}. We have assumed $Z_2$-parity  for the SM singlets to ensure their stability.
  In what follows, we parametrize our results in terms of the physical particle masses $m_s, m_\chi, m_v$, which include both the ``bare'' and the Higgs-portal induced contributions from the above interactions.

 If the dark matter mass exceeds the temperature of the SM thermal bath, its production is Boltzmann-suppressed and, thus,  its coupling to the SM fields can be significant \cite{Cosme:2023xpa}. 
 As a result, this possibility can be probed by current and future direct DM detection experiments as well as by collider searches for invisible Higgs decay.
  
 The abundance of non-thermal dark matter depends on the cosmological history. In the case of freeze-in at stronger coupling, it is sensitive to the 
 temperature evolution. In the following computations,  we make a simplifying assumption of {\it ``instant reheating''}, implying that the SM bath temperature increases quickly from 0 to $T_R$ \cite{Cosme:2023xpa}.
 It gives a good approximation to DM production in models with a flat temperature profile, where $T_{\rm max} \simeq T_R$.
 In this case, the number density is computed by integrating the Boltzmann equation from $T_R$ to $T\sim 0$, assuming a zero initial abundance.
  If the maximal  temperature exceeds $T_R$ but is still well below the DM mass,
 the abundance is obtained  by  replacing $T_R \rightarrow T_{\rm max }$  in all the formulas and      rescaling the result by $(T_R / T_{\rm max})^\alpha$, where $\alpha$ is a model dependent constant \cite{Cosme:2024ndc}.

\section{Scalar dark matter}
 
 Consider spin zero dark matter  $s$ with the Higgs portal coupling $ {1\over 2} \, \lambda_{hs} H^\dagger H \, s^2$.
This is a version of the ``scalar phantom'' dark matter due to Silveira and Zee \cite{Silveira:1985rk}, except in the present  case it never thermalizes and gets produced via Boltzmann-suppressed freeze-in. 
The Higgs coupling provides for a DM production mechanism, so
before we engage in computing the relic DM density, let us consider the issue of Higgs thermalization.

 \subsection{Higgs thermalization}
 
 Since we are interested in low temperatures, the question that needs to be addressed is whether the Higgs boson can be considered a thermal particle at $T\sim\;$GeV. 
 Light fermions $f$  are certainly thermalized at such temperatures due to gauge interactions. Hence, consider the reaction
 \begin{equation}
  \bar f f \leftrightarrow h \;,
 \end{equation}
 which can keep the Higgs boson in thermal equilibrium as long as it is fast enough.
 The Boltzmann equation governing the Higgs number density   ($n_h$)  evolution due to this reaction is
\begin{equation}
\dot n_h + 3H n_h = \Gamma_{\bar f f \rightarrow h} - \Gamma_{h \rightarrow \bar f f } \;,
\end{equation}
where the right hand side (RHS) represents the reaction rates per unit volume and $H$ is the Hubble rate.
In thermal equilibrium,
\begin{equation}
 3H n_h \lesssim  \Gamma_{\bar f f \rightarrow h} \;,  \Gamma_{h \rightarrow \bar f f } \;,
 \label{3Hn}
\end{equation}
so that the reactions are faster than the expansion.

 Consider the Higgs production mode.
The reaction rate is given by 
\begin{equation}
\Gamma_{2\rightarrow 1} = \int \left( \prod_{i} {d^3 {\bf p}_i \over (2 \pi)^3 2E_{i}} f(p_i)\right)~
  {d^3 {\bf p}_f \over (2 \pi)^3 2E_{f}}  
\vert {\cal M}_{2\rightarrow 1} \vert^2 ~ (2\pi)^4 \delta^{(4)}\left(\sum p_i-p_f \right) . 
\label{Gamma}
\end{equation}
Here $p_i$ and $p_f$ are the initial and final state momenta, respectively.
${\cal M}_{2\rightarrow 1}$ is the  QFT  $a \rightarrow b$  transition amplitude. $f(p)$ is the momentum distribution function  for the fermions, which can be taken to be Maxwell-Boltzmann since only the 
particles at the Boltzmann tail   ($E \gg T$)  contribute. 
By energy conservation $f(p_1) f(p_2)= e^{-E_f/T}  $ and $\vert {\cal M}_{2\rightarrow 1} \vert^2  \sim y_f^2 m_h^2 $, where $y_f$ is the fermion Yukawa coupling. The phase space integral reduces to the integral over the final state energy, $ \int \sqrt{E^2-m_h^2} \,e^{-E/T} dE$, which at $m_h \gg T$ reduces to the Gamma function. This yields
\begin{equation}
\Gamma_{2\rightarrow 1} \simeq  {\rm c} \times y_f^2\, m_h^{5/2} T^{3/2}\, e^{-m_h/T} \, ,
\end{equation}
where the constant c$\,\sim 10^{-2}$ includes the spin-color multiplicity factors. Factorizing out the Maxwell-Boltzmann number density 
\begin{equation}
 n_h = \left({m_h T \over 2\pi}\right)^{3/2} e^{-m_h /T} \;, 
\end{equation}
one finds that (\ref{3Hn}) implies
\begin{equation}
y_f \sqrt{m_h M_{\rm Pl}}  \gtrsim T \;,
\label{therm}
\end{equation}
omitting insignificant factors. The same inequality applies to the $1\rightarrow 2$ reaction, 
whose calculation only requires the QFT decay width
$
\Gamma_{\rm QFT} (h \rightarrow \bar f f ) = N_c {m_h y_f^2 \over 8 \pi} \,.
$
The above inequality is easily satisfied at temperatures above the QCD phase transition, hence the Higgs field is always thermalized for our purposes.
A similar statement applies to the gauge bosons $W$ and $Z$.

\subsection{Boltzmann equation and simple estimates}

\begin{figure}[t!]
    \centering
    \includegraphics[width=0.16\textwidth]{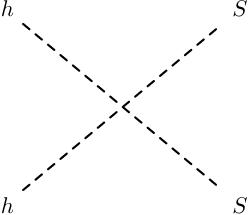} ~~~
     \includegraphics[width=0.135\textwidth]{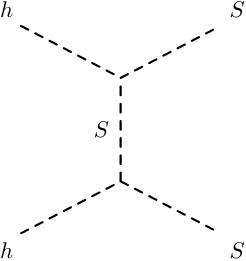}~~~
      \includegraphics[width=0.156\textwidth]{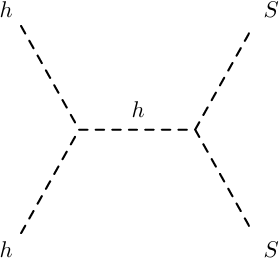}~~~
       \includegraphics[width=0.19\textwidth]{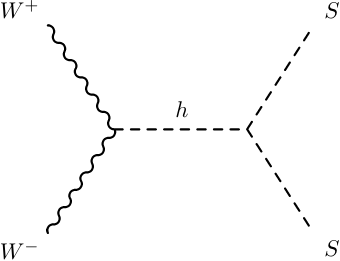}~~~
       \includegraphics[width=0.194\textwidth]{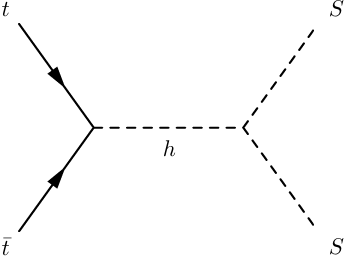}
    \caption{ Leading diagrams for heavy  dark matter production.}
    \label{diags}
\end{figure}

Given that the Higgs boson is in thermal equilibrium in the entire temperature range of interest, we can compute the DM relic abundance due to the Higgs coupling.
  The DM number density obeys the Boltzmann equation
 \begin{equation}
\dot n + 3H n =  2\,\Gamma ({\rm SM} \rightarrow ss) - 2\,\Gamma (ss \rightarrow {\rm SM})   \;,
\label{B}
  \end{equation}
  where $\Gamma$ is the reaction rate per unit volume and ``SM'' denotes collectively the  Standard Model states. 
  The factor of 2 signifies production or annihilation of 2 identical particles.
  For heavy dark matter, the most important channels are pair annihilation of
   the Higgs and  gauge bosons as well as of the top quarks (Fig.\,\ref{diags}).
    When DM is lighter than  half the Higgs mass, it is produced in Higgs decay and annihilation of the light fermions. A  discussion  of the model in the freeze-out and freeze-in regimes can be found in \cite{Burgess:2000yq,Yaguna:2011qn,Lebedev:2019ton}.

    It is instructive to consider the heavy DM limit $m_s^2 \gg m_h^2$.
  Since the temperature of the SM bath $T$ is far below $m_s$, only the particles at the Boltzmann tail, $E/T \gg 1$,    have enough energy for DM pair production. Therefore, the effects of quantum statistics 
 can safely be neglected. The contributions of the gauge bosons can be accounted for by including 4 Higgs degrees of freedom (d.o.f.).
 The    $h_i h_i \rightarrow ss$     reaction rate is given by \cite{Gondolo:1990dk}
 \begin{eqnarray}
  \Gamma (h_i h_i \rightarrow ss)&=& \langle \tilde \sigma(h_i h_i \rightarrow ss)\,   v_r \rangle  \, n_h^2 = {1\over (2\pi)^6} \;     \int \tilde \sigma v_r \, e^{-E_1/T} e^{-E_2/T}  d^3 p_1 d^3 p_2  \nonumber \\
  &=&  {2\pi^2 T\over (2\pi)^6}\;
 \int_{4m_s^2}^\infty  d{\rm s} \; \tilde \sigma \;({\rm s}- 4 m_h^2) \sqrt{{\rm s}} \, K_1 (\sqrt{{\rm s}}/T) \;,
 \label{Gammahhss}
 \end{eqnarray}
 where $\tilde \sigma$ is the   $h_i h_i \rightarrow ss$    cross section which includes the symmetry factor of the initial state 1/2 and the tilde serves to distinguish it from the standard QFT cross section; $v_r$ is the relative velocity of the colliding quanta with energies $E_1,E_2$ and momenta $p_1,p_2$; $n_h$ is the Higgs boson number density;
 $\langle ... \rangle $ denotes a thermal average;
 ${\rm s}$ is the Mandelstam variable, and
 $K_1(x)$ is the modified Bessel function of the first kind.   
 The leading contribution to the cross section is given by the contact term  $s^2h^2$ and 
 \begin{equation}
\tilde \sigma \simeq  4 \times {1\over 2}\times  {\lambda_{hs}^2 \over 32\, \pi {\rm s}  } \; {  \sqrt{{\rm s} - 4m_s^2}   \over   \sqrt{ {\rm s} - 4m_h^2} } \;,
\end{equation}
where we have included the Higgs multiplicity factor $4$ directly in $\tilde \sigma$ (note the definition difference from \cite{Cosme:2023xpa}).  In our convention, we also include the {\it initial and final} state symmetry factors in the cross section.
 Performing the integral at $m_s \gg T$, we get 
 \begin{equation}
  \Gamma   (h_i h_i \rightarrow ss)   \simeq  {1\over 2} \times {\lambda_{hs}^2 T^3  m_s \over 2^7 \pi^4} \,   e^{-2m_s/T} \;,
    \end{equation}
 for four Higgs degrees of freedom. The reaction rate drops exponentially with temperature, as expected.
 
Using
 $H = \sqrt{ \pi^2 g_* \over 90}  {T^2 \over  M_{\rm Pl}}    $ with $g_* $ being the number of the SM effective d.o.f., the Boltzmann equation can be integrated analytically in terms of elementary functions. 
 Starting with zero DM density at $T_R$, the eventual 
   DM abundance    $Y\equiv  n/s_{\rm SM}$    is given by 
 \begin{equation}
Y= {\sqrt{90} \, 45 \over  2^{9} \pi^7 \, g_*^{3/2}  }           \, {\lambda_{hs}^2 M_{\rm Pl} \over  T_R} \,e^{-2m_s/T_R}   \;,
\label{Y1}
\end{equation}
with 4 effective Higgs d.o.f. 
 Imposing the observational   constraint $Y_{\rm obs}= 4.4 \times 10^{-10} \; \left(   {{\rm GeV}\over m_s}  \right) \;,$
 we obtain 
 \begin{equation}
\lambda_{hs} \simeq 3 \times10^{-11}\;  e^{m_s/T_R} \, \sqrt{T_R \over m_s} \;,
\label{result0}
 \end{equation}
for  $g_* \simeq 107$.
This result implies that  $\lambda_{hs} $ has to be tiny unless $m_s/T_R \gg 1$. 
In conventional freeze-in models, $T_R$ is assumed to be above $m_s$, which results in a feeble coupling.
We, on the other hand, are interested in the regime $m_s/T_R \gg 1$.
 For $m_s/T_R \sim 20$, the coupling can be order one. As a result, freeze-in dark matter may be observed, first and foremost, in direct DM detection experiments.

 At larger couplings, the density of produced particles becomes substantial  making the 
 DM annihilation effects significant.
 These are 
 accounted for by
 \begin{equation} 
  \Gamma (ss \rightarrow h_ih_i)  =  \tilde \sigma (ss \rightarrow h_ih_i ) v_r \;n^2~~, ~~ \tilde \sigma (ss \rightarrow h_ih_i ) v_r=  4 \times {1\over 2} \times  {\lambda_{hs}^2 \over {64 \pi m_s^2} }
  \end{equation}
for $m_s^2 \gg m_h^2$ and 4 Higgs d.o.f.
As before, this includes 
 the factor 1/2 due to identical particles in the $initial$ state. The $s$-quanta are non-relativistic, hence we have taken  the zero velocity limit in the above expression.
If the coupling is sufficiently large, the annihilation process  becomes  so efficient that it  leads to thermalization of the $s-h_i$ system. In this case, the production and annihilation rates track each other
for an extended period of time until freeze-out  \cite{Cosme:2023xpa} (see also \cite{Silva-Malpartida:2023yks}).

\subsection{Numerical analysis}

In order to perform the numerical analysis, we rely on the {\tt micrOMEGAs} package \cite{Belanger:2018ccd,Alguero:2023zol} which allows us to efficiently include all possible production channels in our computations. The implementation of 
interactions (\ref{h-s})-(\ref{h-v})  in {\tt micrOMEGAs} has been performed through the {\tt Feynrules} package \cite{Alloul:2013bka}.
\\
\\
In our numerical analysis, 
we use the following version of the Boltzmann equation,
\begin{equation}
\dot n + 3Hn = 2\, \langle \sigma \left(ss\rightarrow {\rm SM}\right)    v_r    \rangle \, \left( n_{\rm eq}^2 -n^2 \right)\;,
\end{equation}
where $\langle ... \rangle$ is the thermal average, $n_{\rm eq}$ is the equilibrium number density and ``SM'' includes all SM final states.
The term $2\, \langle \sigma \left(ss\rightarrow {\rm SM}\right)    v_r    \rangle \, n_{\rm eq}^2$ is responsible for DM $production$ and does not assume any specific distribution
for the DM momenta. Using the Boltzmann distribution and energy conservation, it reduces to $2\,\Gamma ({\rm SM} \rightarrow ss)$ with thermal SM states.
On the other hand, 
the annihilation term $-2\, \langle \sigma \left(ss\rightarrow {\rm SM}\right)    v_r    \rangle \, n^2 $
assumes kinetic equilibrium of DM with the SM bath. This is justified since, due to $n_h \gg n_{ }$, kinetic equilibrium is reached before the annihilation process
becomes important.  A more detailed discussion of the above equation in the context of freeze-in can be found in \cite{Koivunen:2024vhr} (Sec.\,3.1.1).

    \begin{figure}[t!] 
\centering{
\includegraphics[scale=0.95]{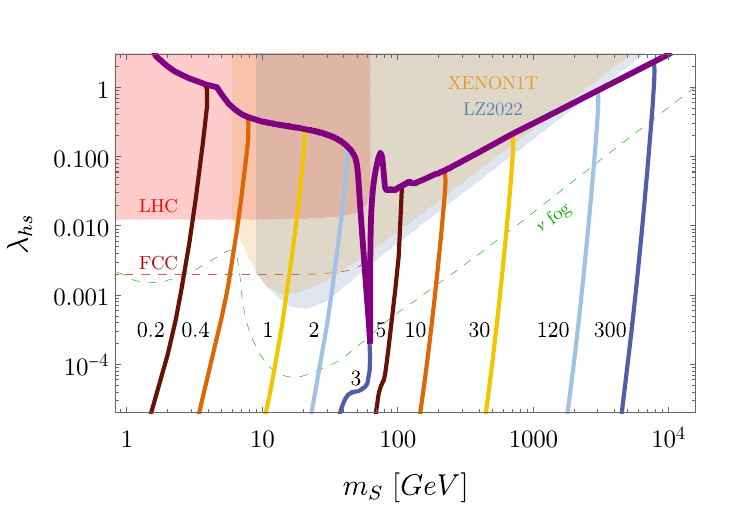}
}
\caption{ \label{par-space-s} {\small
Parameter space of the Higgs portal scalar DM. Along the curves, the correct DM relic abundance is reproduced. The curves are marked by the reheating temperature in GeV, while the purple line corresponds to thermal DM. The shaded areas are  excluded by
the direct DM detection experiments    and the invisible Higgs decay at the LHC. The neutrino background for direct DM detection is represented by the green dashed line ``$\nu$ fog'', while the FCC
prospects  are shown by the red dashed line.}
 }
\end{figure}

\begin{figure}[t!]
    \centering
    \includegraphics[width=0.27\textwidth]{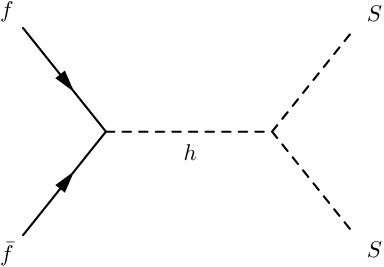} 
          \caption{ Leading  channel for  light   dark matter production.}
    \label{diags1}
\end{figure}

Our results are presented in Fig.\,\ref{par-space-s}. Along the colored lines the correct DM relic density is reproduced for a fixed $T_R$. 
At sufficiently large coupling, 
all the freeze-in lines merge with the thermal (purple) DM curve, where the relic abundance is determined by freeze-out.
In this case, the dependence on $T_R$ is lost. We observe that freeze-in at stronger coupling is consistent with all the constraints in a wide range of masses, from GeV to multi-TeV and above.
Let us consider separately the different DM mass regimes.
\\ \ \\
{\bf \underline{Heavy dark matter}.}
The main contribution to DM production is given by the Higgs and gauge boson initial states. In this case, the approximation (\ref{result0}) 
is adequate and the coupling depends on the DM mass exponentially.  The most important  constraint is imposed by the direct detection experiments  LZ \cite{LZ:2022lsv}, XENON1T \cite{XENON:2018voc} and XENONnT \cite{XENON:2023cxc}.
While LZ is somewhat more sensitive to DM masses above 10 GeV, XENON  provides us with stronger bounds in the range 6 to 10 GeV, with XENON1T  and XENONnT exhibiting similar sensitivity at the moment. 
The DM-nucleon scattering cross section is given by
\begin{equation}
\sigma_{sN} \simeq {\lambda_{hs}^2 f_N^2 \over 4\pi  } \, {m_N^4 \over  m_h^4 \,m_s^2} \;,
\end{equation}
where $f_N \simeq 0.3$ and $m_N\simeq 1\,$GeV.  We observe that DM masses up to 3 TeV are already being probed by direct DM detection.\footnote{Direct DM detection can also be relevant to more complicated freeze-in models, e.g. milli-charged particles with a light mediator  \cite{Hambye:2018dpi}.} As the sensitivity of these searches improves, 
models with a higher $T_R$ and/or lower couplings will continuously be explored. The experiments XENONnT \cite{XENON:2020kmp} and DARWIN \cite{DARWIN:2016hyl} plan to reach the sensitivity levels only bounded by the ``neutrino fog''
background \cite{Billard:2021uyg}, which poses a significant challenge to traditional direct detection experiments.
  \\ \ \\
{\bf \underline{Intermediate and light dark matter}.}
DM production is dominated by light fermions annihilation (Fig.\,\ref{diags1}) and Higgs decay. Close to the Higgs resonance region, $m_s \simeq m_h/2$, 
the Higgs decay $h\rightarrow ss$ plays a significant role.\footnote{At the resonance, the standard thermal  DM abundance calculations receive corrections due to the early ``kinetic decoupling'' of DM
\cite{Binder:2017rgn,Ala-Mattinen:2022nuj}. In our case, the DM phase space distribution is unimportant (away from the thermal DM parameter region)
and this effect is insignificant.}
  This is visible in Fig.\,\ref{par-space-s} at $T_R=3\,$GeV.
The corresponding relic abundance curve has an almost flat patch, which signifies that the abundance is controlled by (the square root of)
the Higgs boson density
 $e^{-m_h/T}$ 
rather than $e^{-m_s/T}$. In most of the parameter space, however, the light fermion annihilation     $\bar f f \rightarrow ss$,
with $m_f < m_s$, is responsible for DM production. This is because the corresponding reaction rate is determined by $e^{-2m_s/T}$ instead of $e^{-m_h/T}$, which controls DM production via Higgs decay. 
The smallness of the Yukawa couplings does not play a significant role in comparison with the exponential suppression. 
As the DM mass decreases, fewer and fewer fermions contribute since their thermal abundance is suppressed by $e^{-m_f/T}$. 
 
 We take the lowest $T_R$ to be 200 MeV to avoid complications with the QCD phase transition. Although lower reheating temperatures are consistent with cosmology, the 
 corresponding computations suffer from significant hadronic uncertainties. Already at $T_R \sim  200\,$MeV, the uncertainty in the effective number of d.o.f. is about 30\%.\footnote{We thank Alexander Pukhov for useful communication concerning hadronic uncertainties at low reheating temperatures.} 
 The very low $T_R$ regime requires a separate analysis, which we reserve for future work.

At $m_s < 6\,$GeV, the most stringent constraint on the model comes from the invisible Higgs decay measurements at the LHC. The corresponding partial width is 
\begin{equation}
\Gamma(h \rightarrow ss)= {\lambda_{hs}^2 v^2\over 32 \pi m_h} \sqrt{1-{4m_s^2 \over m_h^2}} \;.
\end{equation}
In Fig.\,\ref{par-space-s}, we take the Higgs invisible decay branching ratio  ${\rm BR_{inv}} = \Gamma(h \rightarrow ss)/ \left[   \Gamma_{\rm SM} +    \Gamma(h \rightarrow ss) \right]$
to be below 0.11 \cite{ATLAS:2023tkt}. The figure shows that the LHC rules out a significant portion of the parameter space, yet allowing for an observable effect in the pure freeze-in regime. 
Indeed, both the HL-LHC and FCC can probe ${\rm BR_{inv}} $ at the percent/sub-percent level\footnote{In Fig.\,\ref{par-space-s}, we take the projected FCC bound to be 
${\rm BR_{inv}} < 0.003 $ \cite{FCC}. See \cite{Cerri:2016bew} for an earlier study of the Higgs invisible decay at FCC.}, thus exploring the coupling range $\lambda_{hs} \sim 10^{-3}-10^{-2}$. 
The relevance of this observable to low-$T$ freeze-in models was also observed in \cite{Bringmann:2021sth}, while a more general discussion of the invisible Higgs decay into the hidden sector
states can be found in \cite{Biekotter:2022ckj}. More sophisticated freeze-in DM models which involve mediators, can have  complex collider signatures, beyond invisible Higgs decay \cite{Brooijmans:2020yij}.

    We note that the 
indirect DM  detection constraint  in Higgs portal models is superseded by those from direct detection experiments and the LHC \cite{Cline:2013gha}. Hence we omit it in the scalar DM case
(see, however, the fermion DM case below).

\begin{figure}[t!] 
\centering{
\includegraphics[scale=0.57]{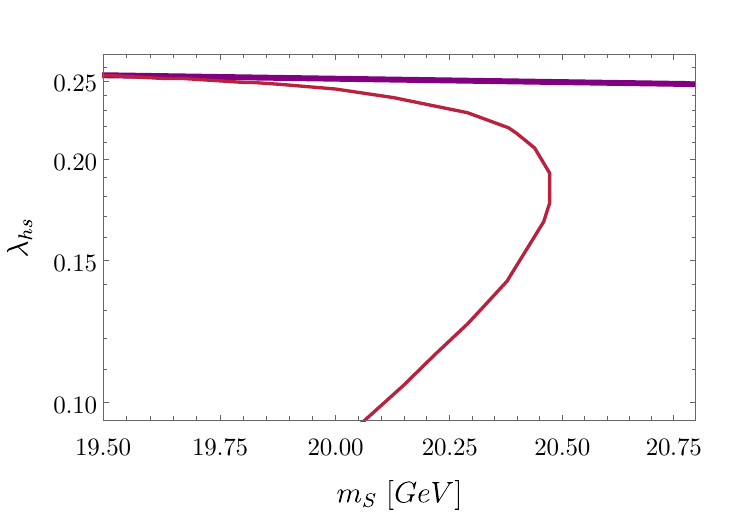}
\includegraphics[scale=0.63]{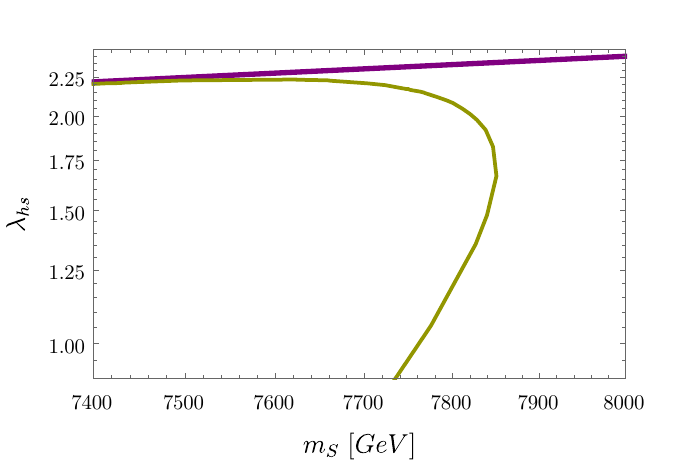}
}
\caption{ \label{FI-FO} {\small
Freeze-in to freeze-out transition at  low and high temperatures. The purple line corresponds to    thermal DM
as in Fig.\,\ref{par-space-s}.
{\it Left:} $T_R=1\;$GeV. {\it Right:} $T_R=300\;$GeV.}
 }
\end{figure}

Finally, it is instructive to examine how the freeze-in lines merge with the thermal abundance curve in Fig.\,\ref{par-space-s}.
 The two panels of Fig.\,\ref{FI-FO}
display the zoomed-in transition areas for $T_R=1\;$GeV and  $T_R=300\;$GeV. We observe that the transition from 
freeze-in  to freeze-out is smooth.
While at low couplings the mass dependence  is exponential, for larger couplings the annihilation channel becomes important and leads to a drastic change in the $\lambda_{hs} (m_s)$ dependence.
As a result, the system thermalizes and the abundance is determined by the usual freeze-out mechanism at $T\sim m_s/20  \lesssim T_R$.

\section{Fermion dark matter}

The fermion dark matter option \cite{Kim:2006af}   differs significantly from the scalar case. Since the coupling is effective, the model is meaningful up to the cutoff  of order $\Lambda$ or $\Lambda_5$. The typical energy of the relevant processes is about $2 m_\chi$,
so the cutoff must be above this value and, in addition, above the electroweak scale. Unlike the scalar case, the fermion DM phenomenology exhibits velocity dependence, which affects the constraints and possible signatures.  
These differ significantly in the CP-conserving     and   (maximally)  CP-violating cases. In what follows, we consider the two options separately.

\subsection{CP-conserving coupling}

This corresponds to the Higgs portal interaction
$     1/  \Lambda \; H^\dagger H \, \bar \chi   \chi    \,$,
where $\chi$ is   a Majorana fermion. The relevant DM production/annihilation  diagrams are similar to those in the scalar case. The analysis proceeds analogously except the spin structures bring in additional momentum dependence.

It is instructive to consider again the heavy dark matter case in the pure freeze-in limit, which allows for simple analytical estimates.
The DM production cross section in the heavy mass limit is 
\begin{equation}
\tilde \sigma (h_i h_i \rightarrow \chi \chi) = {1\over 2 \pi \Lambda^2} \, \left( {1- {4m _\chi \over {\rm s}}}  \right)^{3/2}\;,
\end{equation}
where we include 4 Higgs d.o.f. as well as 1/2 for identical particles in the initial state.
\\
\\
The corresponding reaction rate per unit volume is
\begin{equation}
\Gamma (h_i h_i \rightarrow \chi \chi) \simeq {3\over 2^4 \pi^4} {m_\chi^2 T^4 \over \Lambda^2}\, e^{-2 m_\chi /T} \;.
\end{equation}
We note that it scales with the temperature as $T^4 e^{-2 m_\chi /T}$ compared to $T^3 e^{-2 m_\chi /T}$ in the scalar case. This is due to the stronger velocity dependence of the corresponding cross section.
The resulting relic abundance is found by solving  $\dot n + 3Hn =2 \Gamma$ and is given by 
\begin{equation}
Y \simeq 1.2 \times 10^{-5} \; {m_\chi M_{\rm Pl} \over \Lambda^2 } \, e^{-2 m_\chi /T_R}  \;,
\end{equation}
which implies
\begin{equation}
  {m_\chi \over \Lambda} \, e^{- m_\chi /T_R}  \simeq 4 \times 10^{-12}\;.
\end{equation}
If $m_\chi$ and $\Lambda$ are not vastly different, this requires $m_\chi /T_R \sim 26$.

    \begin{figure}[t!] 
\centering{
\includegraphics[scale=0.65]{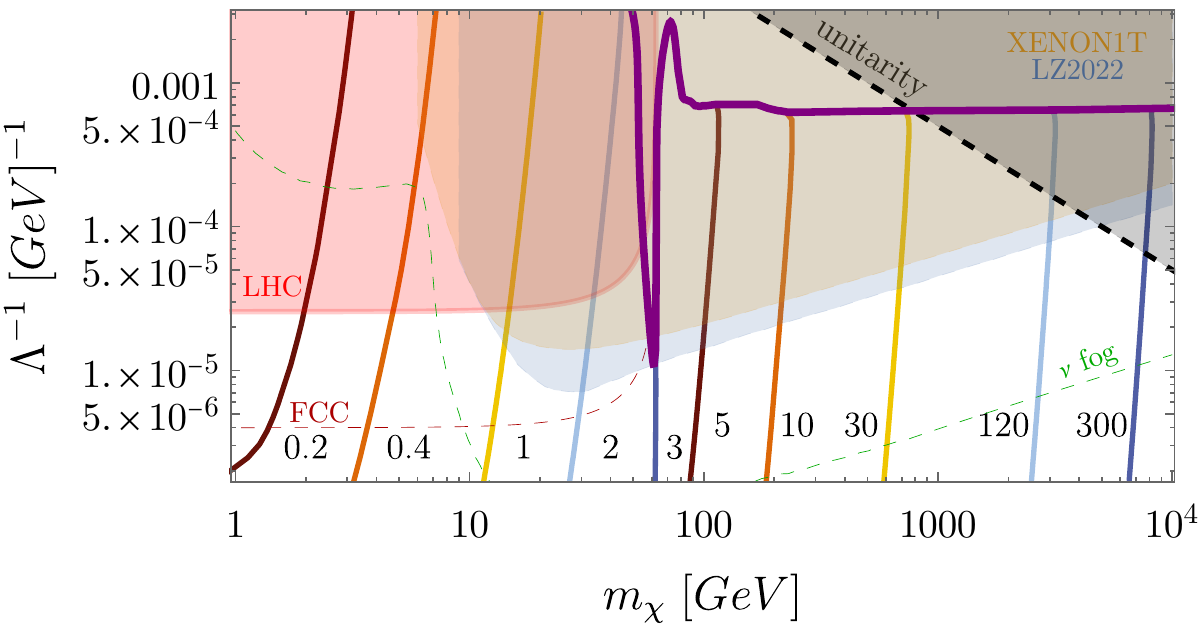}
}
\caption{ \label{par-space-f1} {\small
Parameter space of the Higgs portal fermion DM with a CP-even coupling. The curves are labeled by $T_R$ in GeV. (See Fig.\,\ref{par-space-s} for details.) The unitarity constraint is defined by  $\Lambda > 2 m_\chi$.} 
 }
\end{figure}

The most important constraints  on the model are imposed by the direct detection bounds and the Higgs decay data, the latter being relevant for light DM only.
The invisible Higgs decay  into Majorana fermions has the width
\begin{equation}
\Gamma (h \rightarrow \chi \chi)= {m_h \over 4 \pi} \, {v^2 \over \Lambda^2} \, \left(1- {4 m_\chi^2 \over m_h^2} \right)^{3/2} \;.
\end{equation}
The spin-independent DM-nucleon cross section is 
\begin{equation}
\sigma_{\chi N} = {4\over \pi \Lambda^2 m_h^4} \, { m_N^4 m_\chi^2 \over (m_N + m_\chi)^2} \, f_N^2 \;,
\end{equation}
where $f_N \simeq 0.3$. Our numerical results and the constraints are shown in Fig.\,\ref{par-space-f1}. The curve legends are the same as those in the scalar case (Fig.\,\ref{par-space-s}). 
We observe substantial  similarities between the scalar and CP-even fermionic DM options, except in the latter case the direct detection constraint excludes much larger regions of parameter space.   

It is well known that thermal fermionic Higgs portal DM is essentially excluded \cite{Djouadi:2011aa}, apart from the resonance region \cite{GAMBIT:2018eea}. The annihilation process is impeded by the velocity suppression, while the direct detection amplitude is unsuppressed. This makes the relic abundance and direct detection constraints incompatible as seen in Fig.\,\ref{par-space-f1}. Note that the curve corresponding to the thermal abundance becomes flat for heavy DM, which is due to the fact that $\langle \sigma_{\rm ann} v \rangle$ depends on the velocity ($\sim 1/20$) and $\Lambda$ only, being  independent of the DM mass. The low mass region requires a large coupling and is excluded by the invisible Higgs decay. We note that the latter imposes a stronger constraint than the direct detection could potentially achieve
since the corresponding bound lies below the ``neutrino floor'' line.

In the freeze-in case, however, such problems do not arise and the model is viable at weak couplings. In this regime, the annihilation effects are negligible corresponding to pure freeze-in. 
At $m_\chi \sim 1\,$GeV, we see substantial  curving of the relic abundance lines. This is because of the  significant decrease   in the size of the relevant Yukawa couplings: 
the tau and charm quarks ``decouple'' due to the exponential suppression of their abundance, such that  DM is mostly produced by the lighter strange quarks.
  In terms of observational prospects, 
 we see that, as in the scalar case,  the model can be probed both by the direct DM detection experiments and the collider searches for the invisible Higgs decay.
The former is largely limited by the neutrino fog, while the latter is unlikely to probe branching ratios below few$\,\times 10^{-3}\, \%$. Yet, there is a large portion of the parameter 
space where the model is viable and can be explored further.  The above analysis and conclusions are analogous to those of \cite{Koivunen:2024vhr}, which studied Boltzmann-suppressed  freeze-in signatures in the singlet
extended SM with sterile neutrinos as dark matter.

\subsection{CP-violating coupling}

\begin{figure}[t!] 
\centering{
\includegraphics[scale=1.09]{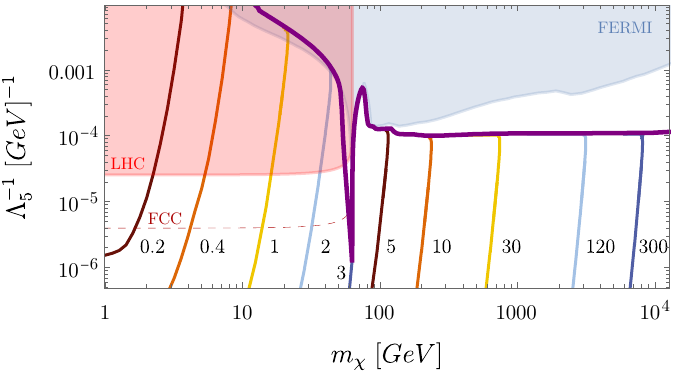}
}
\caption{ \label{par-space-f2} {\small
Parameter space of the Higgs portal fermion DM with a CP-odd coupling. The curves are labeled by $T_R$ in GeV. (See Fig.\,\ref{par-space-s} for details.)
The direct DM detection bounds are weak and superseded by the indirect DM detection constraint from FERMI.}
 }
\end{figure}

The Higgs portal interaction is now $     1/  \Lambda_5 \; H^\dagger H \, \bar \chi  i \gamma_5 \chi    $, leading to vastly different phenomenology \cite{Lopez-Honorez:2012tov}.
  The DM production cross section in the heavy mass limit is 
\begin{equation}
\tilde \sigma (h_i h_i \rightarrow \chi \chi) = {1\over 2 \pi \Lambda^2_5} \, \sqrt{1- {4m _\chi \over {\rm s}}}  \;,
\end{equation}
where we include 4 Higgs d.o.f. as well as 1/2 for identical particles in the initial state. As in the scalar case, this asymptotic behaviour is determined by the contact term.
The corresponding reaction rate per unit volume is
\begin{equation}
\Gamma (h_i h_i \rightarrow \chi \chi) \simeq {1\over 2^3 \pi^4} {m_\chi^3 T^3 \over \Lambda^2_5}\, e^{-2 m_\chi /T} \;.
\end{equation}
The temperature dependence is the same as in the scalar case, $T^3 e^{-2 m_\chi /T} $. Solving the Boltzmann equation $\dot n + 3Hn =2 \Gamma$, we get
\begin{equation}
Y \simeq 8 \times 10^{-6} \; {m_\chi^2 M_{\rm Pl} \over \Lambda^2_5 T_R} \, e^{-2 m_\chi /T_R}  \;,
\end{equation}
which implies that the correct relic DM abundance requires
\begin{equation}
{m_\chi^{3/2}  \over \Lambda_5 \, T_R^{1/2} } \, e^{- m_\chi /T_R}  \simeq 5\times 10^{-12} \;,
\end{equation}
in the heavy DM limit. The full numerical results are displayed in Fig.\,\ref{par-space-f2}.

At large enough couplings, the annihilation term becomes important and DM thermalizes. 
The DM abundance is then determined by the usual freeze-out.
In the heavy mass limit, the relic abundance fixes the coupling independently of the mass (Fig.\,\ref{par-space-f2}, purple curve). 
 Indeed, the combination
$\tilde \sigma (\chi \chi \rightarrow h_i h_i)  v_r \propto 1/(4\pi \Lambda^2_5)$ is constant for heavy DM. Thus, the corresponding DM energy density $\Omega h^2 \propto 1/ \langle \tilde \sigma v_r \rangle$
is mass-independent, which explains the flat thermal abundance line.

The freeze-in DM signatures differ from those in the CP-even case.
Since the direct detection cross section is  highly velocity suppressed by the factor $v_0^2 \sim 10^{-6}$, the relevant constraints are imposed by indirect DM detection and the Higgs decay data.  
The leading indirect detection constraint stems from the Fermi observations of the Milky Way    satellite    dwarf spheroidal galaxies. 
In Fig.\,\ref{par-space-f2}, we show the observed limit \cite{McDaniel:2023bju} based on the  $b\bar{b}$ DM annihilation channel. The gauge boson final states lead to a similar bound.
\\
\\ 
The invisible Higgs decay  into Majorana fermions has the width
\begin{equation}
\Gamma (h \rightarrow \chi \chi)= {m_h \over 4 \pi} \, {v^2 \over \Lambda_5^2} \, \sqrt{1- {4 m_\chi^2 \over m_h^2}} \;.
\end{equation}
The resulting constraint excludes $\Lambda_5$ below about 100 TeV, for $m_\chi < m_h/2$. Indirect DM detection only adds a small segment of the resonance region to the excluded parameter space. Resonant annihilation in the Early Universe and at current times occurs  at different velocities which results in a slight shift of the resonance dip in the indirect detection constraint.

Unlike in the CP-even case, the parameter space at $m_\chi \gtrsim m_h/2$ is essentially unconstrained and the thermal DM option is consistent with the data. Certain low mass regions could be probed to some extent 
by future indirect detection experiments. At larger $m_\chi$, the gap between the current bound and the thermal DM line   in  Fig.\,\ref{par-space-f2}  grows making indirect detection very challenging.
 On the other hand, light DM with $m_\chi \lesssim m_h/2$ can be explored via  collider searches for invisible Higgs decay, similarly to the CP-even case.

  \section{Vector dark matter}

The case of vector dark matter with the coupling ${1\over 2}\, \lambda_{ hv } H^\dagger H \, V_\mu V^\mu $ is largely similar to that of scalar DM. In our framework, dark matter is produced in the deep non-relativistic regime such that a massive vector is equivalent to 3 degenerate scalars.
That is, one may replace $V_\mu V^\mu \rightarrow \sum_i s_i^2$, where $s_{1,2,3}$ represent the different spin projection states. It is then clear that, in the pure freeze-in regime, the DM abundance triples compared to the scalar case
if $\lambda_{hv}=\lambda_{hs}$. Therefore, to obtain the correct relic abundance, one may simply rescale the scalar case result,
\begin{equation}
\lambda_{hv}\Big\vert_{\rm FI} = {1\over \sqrt{3}} \, \lambda_{hs}\Big\vert_{\rm FI} \;,
\end{equation} 
keeping the DM mass and $T_R$ intact.

In the freeze-out regime, this  behaviour reverses. Treating $s_i$ as independent scalars, one finds that each of them freezes out at the same temperature $T_{\rm FO}$ as in the single scalar case. Therefore, the DM abundance triples  
and one has to increase the annihilation cross section by a factor of 3 to obtain  the correct relic density. Hence, the rescaling factor becomes
\begin{equation}
\lambda_{hv}\Big\vert_{\rm FO} = { \sqrt{3}} \, \lambda_{hs}\Big\vert_{\rm FO} \;,
\end{equation} 
for other parameters fixed.

   \begin{figure}[t!] 
\centering{
\includegraphics[scale=0.84]{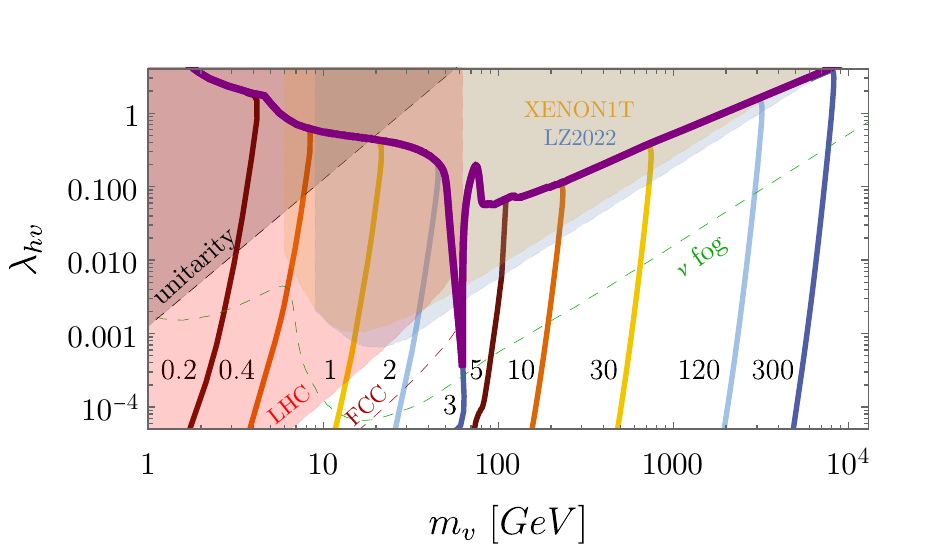}
}
\caption{ \label{par-space-v} {\small
Parameter space of the Higgs portal vector  DM. The curves are labeled by $T_R$ in GeV. (See Fig.\,\ref{par-space-s} for details.)
Above the unitarity constraint line, the effective field theory   description is inadequate. } 
 }
\end{figure}

Our numerical results are presented in Fig.\,\ref{par-space-v}.
While the relic abundance lines shift somewhat compared to the scalar case, the direct detection bound remains the same at $\lambda_{hv}=\lambda_{hs}$:
\begin{equation}
\sigma_{vN} = \sigma_{sN} \;,
\end{equation}
since the three scalars have the same nucleon scattering cross sections. 
The invisible Higgs decay width, on the other hand, increases fast as the vector becomes lighter:
\begin{equation}
\Gamma (h \rightarrow VV)= {\lambda_{hv}^2 v^2 m_h^3 \over 128 \pi m_v^4} \,   \left(    1-4  {m_v^2 \over m_h^2}+12 {m_v^4 \over m_h^4}    \right)   \,\sqrt{1-4 {m_v^2 \over m_h^2}} \;.
\end{equation}
 This growth, however, is bounded by applicability of our effective field theory framework, as we discuss below.

The validity of the vector Higgs portal  description is limited by unitarity considerations \cite{Lebedev:2011iq,Arcadi:2020jqf,Arcadi:2023gox}. 
The corresponding coupling is effective, as required by gauge symmetry, and thus becomes inadequate above a certain energy scale. This scale is proportional to the vector mass so that the effect 
is particularly important for light vectors.
For example, the Higgs-mediated $VV \rightarrow VV$ scattering amplitude grows as $E^4$ with the energy scale  (below $m_h$) and becomes large above $\Lambda \sim m_v /\sqrt{\lambda_{hv}} \,\times\, c$, where $c$ is a constant.
At larger energies ($\gtrsim m_h$), the process $hh \rightarrow VV$ imposes a similar cutoff.
The existence of the cutoff can also be seen  from the perspective of the Higgs mechanism in the dark sector \cite{Lebedev:2011iq}.
 Integrating out the dark sector ``Higgs'' which has a portal coupling to the SM Higgs, one obtains an effective coupling
$\lambda_{hv } \sim \lambda_{hs}\, m_v^2/m_s^2$, where $m_s$ is the mass of the dark Higgs. 
The portal coupling $\lambda_{hs}$ is bounded by perturbativity to  about ${\cal O}(4\pi)$, then requiring $m_s$ to be the cutoff leads to the bound 
similar to the one from unitarity considerations.

 For light dark matter, we require the effective approach to be valid with regard to the Higgs decay calculations. Hence, we impose
 \begin{equation}
 {\rm const } \times {m_v \over \sqrt{\lambda_{hv}}}  \gtrsim 100~{\rm GeV} \;,
 \end{equation} 
 where the constant is of order $\sqrt{4\pi}$.
 For heavy DM, we only make sure that the theory is valid up to the scale $2m_v$. This does not impose any further constraints. Indeed, 
 dark matter is produced by the thermal bath in the deep non-relativistic regime and any heavier states are irrelevant. 
 
  Fig.\,\ref{par-space-v} shows that there are large regions of parameter space at weak coupling with $T_R \sim \,$GeV, where the unitarity constraint is satisfied and the invisible Higgs decay provides one with a sensitive probe of
  vector dark matter. The LHC is already exploring couplings below $10^{-4}$ for $m_v < 10\,$GeV and the FCC will gain a further factor of 6. In the mass range $m_v < 20\,$GeV, the collider bound is stronger than
  the direct DM detection constraint and this situation will persist in the future due to the ``neutrino fog'' limitations.
  
  We note that the relic density calculations are not physically meaningful inside the unitarity-excluded region.  In particular, the Higgs boson decay width explodes making our effective approach inadequate. The relic density curves displayed in Fig.\,\ref{par-space-v}  in that region correspond to a formal rescaling of the analogous scalar DM curves and carry no physical significance.

\section{Conclusion}

We have studied freeze-in dark matter phenomenology in the Higgs portal framework. When the dark matter mass exceeds the SM thermal bath temperature, the DM production is Boltzmann-suppressed and proceeds in the regime of freeze-in at stronger coupling. This mechanism interpolates between   the usual DM freeze-in and freeze-out. It retains the flexibility of the former while keeping the observational signatures of the latter.

The purpose of this work is to delineate allowed parameter space in the simplest Higgs portal models with  dark matter spin 0,1/2, 1. For couplings below those required by the usual freeze-out mechanism, freeze-in dark matter is viable
in a wide range of masses from 1 GeV to 10 TeV. The model properties are largely controlled by the reheating temperature $T_R$. In our parameter space analysis, we resort to  the instant reheating approximation,
which serves a benchmark for assessing the efficiency of DM production. The results  can be applied to a wider variety of reheating scenarios using appropriate rescaling, as long as the DM mass remains  above the SM bath temperature at any stage of the evolution.

The main constraints on the model are imposed by the direct DM detection experiments and collider searches for invisible Higgs decay. These rule out significant regions of freeze-in DM parameter space. On the other hand, 
they also provide one with the avenues to probe the model further. Current and future direct detection experiments such as XENONnT and DARWIN will be able to probe DM-nucleon scattering down to the ``neutrino floor'',
where the background effects become overwhelming. This allows us to explore smaller Higgs portal couplings and/or higher reheating temperatures.
Light dark matter can be searched for at colliders via invisible Higgs decay. The Boltzmann-suppressed freeze-in allows for substantial DM-Higgs couplings such that 
the invisible Higgs decay branching ratio can be at the level of 10\% and below. This range will be probed by the LHC, with smaller (sub-percent)  values accessible at the  FCC.
\\ \ \\
{\bf Acknowledgements.} The work of F.C. is supported by the European Union's Horizon 2020 research and innovation programme under the Marie Sklodowska-Curie grant agreement No 860881- HIDDeN.
This work used the Scientific Compute Cluster at GWDG, the joint data center of Max Planck Society for the Advancement of Science (MPG) and University of G\"ottingen.
A.G. would like to thank G. B\'elanger, F. Boudjema and A. Pukhov for useful discussions.

\end{document}